\documentclass{article}
\usepackage[utf8]{inputenc}

\usepackage[acronym,nomain,nonumberlist]{glossaries}

\newacronym{SC-1}{\mbox{SC-1}}{Standard Cleaning 1}
\newacronym{SiO2}{SiO$_{2}$}{silicon dioxide}
\newacronym{PECVDd}{PECVD}{ plasma-enhanced chemical vapor deposited}
\newacronym{PECVD}{PECVD}{ plasma-enhanced chemical vapor deposition}
\newacronym{NH3}{NH$_{3}$ (aq.)}{ammonia solution}
\newacronym{H2O2}{H$_{2}$O$_{2}$}{hydrogen peroxide}
\newacronym{H2O}{H$_{2}$O}{deionized water}
\newacronym{SEM}{SEM}{scanning electron microscope}
\newacronym{AFM}{AFM}{atomic force microscope}
\newacronym{AFMy}{AFM}{atomic force microscopy}
\newacronym{Nb}{Nb}{niobium}
\newacronym{Si}{Si}{silicon}
\newacronym{TiN}{TiN}{titanium nitride}

\title{SC-1 Etching of Niobium and Titanium Nitride Thin Films}
\author{Adrián Gutiérrez-Cruz*, K. A. C. Rathnathilaka$^{\dagger}$, Jani M. Taskinen, \\
Tuomas Vaimala, Kestutis Grigoras, Harshad Mishra, Rishabh Upadhyay, \\ Jorden Senior, Alberto Ronzani$^{\ddagger}$ \\\\
*adrian.gutierrezcruz@vtt.fi\\
$^{\dagger}$achini.rathnathilaka@vtt.fi\\
$^{\ddagger}$alberto.ronzani@vtt.fi\\
VTT Technical Research Centre of Finland, Tietotie 3, Espoo, Finland}

\date{}
\usepackage[a4paper, total={15cm, 20cm}]{geometry}
\usepackage{graphicx}
\usepackage{mathtools}
\usepackage{multirow}
\usepackage[numbers,square,sort&compress]{natbib}

\begin{document}

\maketitle
\begin{abstract}
Dry etching techniques, ubiquitous in microelectronics fabrication, often result in challenging levels of undesired collateral plasma-induced damage. 
In this work, we demonstrate a wet etching alternative for the patterning of niobium (Nb) and titanium nitride (TiN) thin films using the Standard Cleaning 1 (SC-1) solution. 
We characterize the etching process through its time-evolution dynamics, supported by scanning-electron and atomic force microscopy assessment of the etched film morphology. The results suggest etch dynamics that are linked to native oxides and film microstructure.
Overall, the manageable etch rates, the safe operation and the high material selectivity are attractive for practical use in microelectronics fabrication.
\end{abstract}


\section*{Introduction}
Achieving precise and reliable thin film patterning in fabrication is essential for ensuring the functionality and performance of electronic devices. Dry etching patterning techniques such as reactive ion etching and inductively coupled plasma etching are commonly employed due to their high etch rates and their ability to produce anisotropic profiles, resulting in well-defined features \cite{Dryetch}. However, these techniques can also present certain challenges such as limitations in material selectivity, making it difficult to remove specific materials without affecting other layers \cite{Dryetch1, Dryetch2}, and plasma-induced damage (PID) on substrates due to the highly energetic ion bombardment, which can affect material properties and device performance \cite{PID, PID1}. Therefore, it is important to consider other etching methods that may offer improved results.

In this study, motivated by the challenge of achieving selective etching of \gls{Nb} and \gls{TiN} on top of phosphorus-doped silicon (Si) regions without causing PID on the substrate~\cite{supplementary}, we tested a wet etching technique based on the RCA \gls{SC-1} solution \cite{SC1}, for the patterning of \gls{Nb} and \gls{TiN} thin films. The aim of this work is to document this approach as an alternative to dry etching techniques and expand the portfolio of microfabrication techniques available for these two metals. Commonly, \gls{Nb} and \gls{TiN} are used as metallization layers in electronic devices \cite{metall1, metall2, metall3} due to their established fabrication protocols, compatibility with semiconductor manufacturing, and valuable properties in electronic applications, such as superconductivity \footnote{Both metals have a relatively high superconducting transition temperature: Nb at approximately \mbox{9.2 K} \cite{Nb} and TiN between \mbox{4 to 6 K} \cite{TiN,TiN1}, depending on stoichiometry and film microstructure.}.

The \gls{SC-1} solution was initially developed to clean Si wafers from particles and organic residues \cite{SC1}, but has also been reported to etch different materials \cite{TiN2}. One notable example is \gls{TiN}, in which the etchant has been particularly useful for the fabrication of various electronic devices including transmon qubits and transistors \cite{TiNap1, TiNap2}. However, detailed information about the influence of film microstructure during \gls{SC-1} etching remains limited, making it difficult to predict etching outcomes, which in turn influences fabrication efficiency, reliability, and yield \cite{elec}. Most studies focus on applications rather than its systematic etch behavior, highlighting the need for further investigation \cite{TiN2, TiNap2,TiN3}. In the case of Nb and the use of \gls{SC-1} solution as etchant, there have been initial studies conducted by us \cite {Nbet4}, but research in this area remains scarce. Beyond this specific investigation, to our knowledge, there are no reports detailing a wet etching approach that utilizes this particular etchant for \gls{Nb} \cite{Nbet1, Nbet2, Nbet3}.

\section*{Methods}
The etching process was tested on \mbox{1 cm x 2 cm} sized chip samples, diced from a wafer fabricated as follows: a 150 mm Si wafer (\mbox{675 µm} thick) was first capped with a thermally grown \gls{SiO2} layer of approximately \mbox{200 nm}, serving as substrate. Thin films of either Nb or TiN were subsequently sputtered, with thicknesses of \mbox{58 $\pm$ 3 nm} and \mbox{128 $\pm$ 3 nm}, respectively \footnote{Average thicknesses estimated from sixty-four Nb chips and fifteen TiN chips diced across the wafers. Deposition uniformity was 14\% for Nb and 6\% for TiN.}.

\begin{figure}[ht]
  \centering
  \includegraphics[scale=1.1]{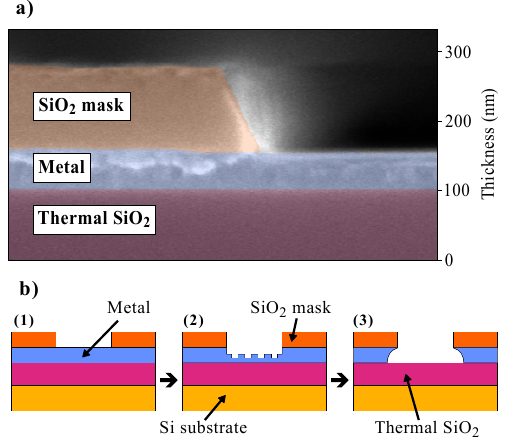}
  \caption{a) Cross-sectional \gls{SEM} image of a chip at the edge of a mask opening prior to wet etching, showing the initial layer stack. b) Schematic of the progressive wet etching of a chip against time: (1) before etching, where the metal film is exposed within the \gls{SiO2} hard mask mask opening, (2) during etching, where an increase in the metal film roughness is observed, (3) after etching, where the metal film within the mask opening is fully etched and showing a potential overetching of the metal side-walls under the mask.}
  \label{Fig1}
\end{figure}

The films were protected with a patterned \gls{PECVD} \gls{SiO2} hard mask (nominal thickness of \mbox{100 nm}). This allowed us to estimate the vertical etched depth, having the masked metal as a non-etched reference. The pattern design comprised a set of reticles with square mask openings. A \mbox{cross-sectional} \gls{SEM} image of a Nb chip before etching is presented in \mbox{Figure \ref{Fig1}a)}, showing the edge of a mask opening where metal is exposed for wet etching process.

The patterning of the \gls{SiO2} hard mask was done differently for each metal. For the Nb film, i-line  \mbox{UV mask} projection lithography and CHF$_{3}$/CF$_{4}$/Ar plasma etching were employed. For the TiN film a laser maskless aligner and CHF$_{3}$/Ar etching approach were used. Due to lack of dry etch selectivity, the \mbox{TiN film} within the patterned mask openings was unintentionally overetched and its thickness was reduced to \mbox{104 $\pm$ 4 nm}, henceforth considered as the initial thickness before wet etching. In contrast, the Nb film was not overetched owing to its reduced sensitivity to the etching chemistry and more precise control on the etching endpoint. After patterning, wafers were diced in individual chips (each one containing 38 patterned mask openings) for the following wet etching. 

The \gls{SC-1} solution was prepared using the typical volume ratio of \mbox{1 part} of ammonia (NH$_{3}$) aqueous solution 25$\%$, \mbox{1 part} of hydrogen peroxide (H$_{2}$O$_{2}$) aqueous solution 30$\%$, and \mbox{5 parts} of \gls{H2O}. The solution was prepared immediately before each experiment, first mixing NH$_{3}$ and \gls{H2O} together, then heating up the solution to the target temperature, and finally adding H$_{2}$O$_{2}$ to it. The reason of this approach is to minimize H$_{2}$O$_{2}$ thermal decomposition over time and prevent drifts on stoichiometry  \cite{H2O2}. The addition of H$_{2}$O$_{2}$ to the solution causes a temperature drop of approximately \mbox{5°C}, thus target temperature has to be reached again before starting with the etching. Furthermore, in view of the fact that H$_{2}$O$_{2}$ decomposes over time, the etching was performed as quickly as possible and within a maximum of 30 min after H$_{2}$O$_{2}$ addition for all the samples. Nb films were etched at \mbox{65°C}, whereas TiN films were etched at \mbox{45°C}.

The chips were immersed in the etchant solution for varying durations.  After reaching the target etching time, chips were soaked in a water bath with ultrasound for approximately one minute to stop the etch reaction and clean the samples from residues. Finally, chips were dried with a N$_{2}$ gun and cleaved across the mask openings to image the \mbox{cross-section} on \gls{SEM} and estimate the etched depth. Figure \ref{Fig1}b) shows a depiction of the etching process and \mbox{Table \ref{Tab1}} summarizes the process parameters and calculated etch rates.

\begin{table}[ht]
\centering
\begin{tabular}{|c|c|c|c|}
\hline
Material & Etchant           & Temperature (°C)        & Etch rate (nm/min) \\ \hline
     Nb &  \multirow{2}{*}{Standard Cleaning 1 solution} & 65 & 50 \\ \cline{1-1} \cline{3-4} 
      $\mathrm{NbO_{x}}$* & & 65 & 8 \\ \cline{1-1} \cline{3-4} 
      TiN & (1 NH$_{3}$ : 1 H$_{2}$O$_{2}$ : 5 H$_{2}$O) & 45 & 6 - 32$^{\dagger}$ \\ \hline
\end{tabular}
\caption{Process parameters and estimated etch rates. NH$_{3}$ = Ammonia aqueous solution 25\% ; H$_{2}$O$_{2}$ = Hydrogen peroxide aqueous solution 30\% ; H$_{2}$O = Distilled water.\\ *Thin native oxide, present on the Nb surface \\ $^{\dagger}$Multi-modal etching behaviour, see text and Figure \ref{Fig2}.}
\label{Tab1}
\end{table}

\section*{Results and Discussion}
\begin{figure}[ht]
  \centering
  \includegraphics[scale=1.1]{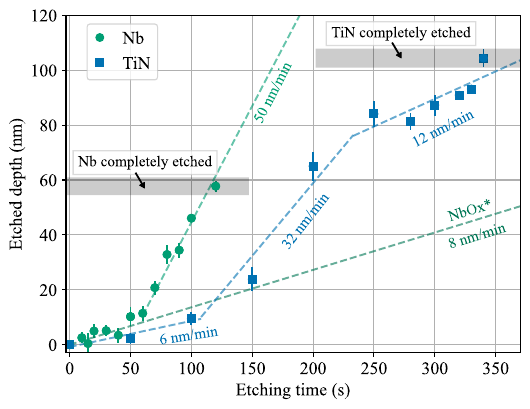}
  \caption{Etched depths estimated by cross-sectional \gls{SEM} imaging as a function of etching time for the Nb (green circles) and TiN (blue squares) films. The grey-shaded horizontal areas show each film’s etching endpoint based on its estimated initial thickness.\\ *The 8 nm/min rate is attributed to the NbO$_{\text{x}}$ etching, considering the interpretation given in text.}
  \label{Fig2}
\end{figure}

Figure \ref{Fig2} shows the etched depths estimated by cross-sectional \gls{SEM} imaging as a function of etch time, for both Nb and TiN films. The etching of the Nb film (green circles) results in a bimodal process, composed of a slow and fast etching phase. Considering that Nb naturally grows an oxide layer when exposed to air \cite{NbO1,NbO2}, we speculate that the initial slow etching phase observed is related to the presence of this native oxide layer (NbO$_{\text{x}}$). According to previous reports, this layer is composed of a mixture of NbO, NbO$_{2}$, and Nb$_{2}$O$_{5}$ and has a thickness of \mbox{5-7 nm} after a self-limiting time of approximately 10$^{3}$ h \cite{NbO2,NbO3}.

After the NbO$_{\text{x}}$ film has been removed, the etching process continues to the Nb, exhibiting a higher etch rate compared to its oxides. Under this interpretation, calculated etch rates are \mbox{8 nm/min} for NbO$_{\text{x}}$ and \mbox{50 nm/min} for Nb. We also note that the etching selectivity of \gls{SC-1} with respect to both the thermally-grown and \gls{PECVD} \gls{SiO2} layers is high, as no detectable etching effect is observed even under prolonged exposure\footnote{Considering 
the uncertainty of the initial \gls{SiO2} mask thickness estimates (8~nm) and the maximum etching time tested of 600~s, we estimate the \gls{SiO2} etch rate to be less than 1~nm/min.}.

The TiN film (blue squares in Figure \ref{Fig2}), on the other hand, exhibits a multi-modal etching process, composed of an initial rate of \mbox{6 nm/min}, followed by a temporary increase in the etching rate up to \mbox{32 nm/min}, and a final phase with an approximate rate of \mbox{12 nm/min}. Different from Nb, TiN films do not develop a native oxide in ambient conditions, due to its chemical stability \cite{TINOx1, TiNox2, TiNox3}, hence, a different origin for the non-linear behaviour must be suspected.

\begin{figure}[ht]
  \centering
  \includegraphics[scale=0.9]{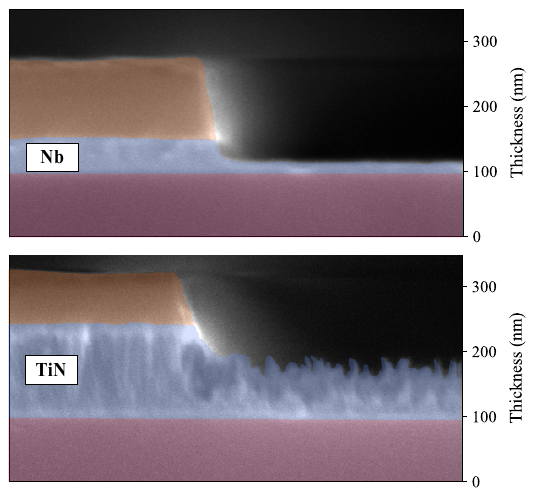}
  \caption{Cross-sectional \gls{SEM} images of the etched area from the Nb and TiN films when the etching process is approximately half-way: after 80 s and 150 s for Nb and TiN, respectively.}
  \label{Fig3}
\end{figure}

Figure \ref{Fig3} illustrates cross-sectional SEM images of the Nb and TiN films after \mbox{80 s} and \mbox{150 s} of etching, respectively. The TiN film shows a pronounced columnar structure, coarser grain size, and a rougher surface at the etched area compared to the Nb film.  The columnar structure observed is common for transition metal nitrides \cite{Nitrides} and especially for TiN due to the competitive growth mechanism between the \mbox{TiN(111)} and \mbox{TiN(200)} crystal plane orientations during the sputtering deposition \cite{TiN_sputtering1,TiN_sputtering2,TiN_sputtering3}.

Morphological characteristics affect the etching dynamics of films, in particular when different crystal orientations have different etch rates, resulting in a vertically non-uniform etching process \cite{Cu}. This mechanism ultimately leads to an increase in film roughness and effective surface area, which can in turn, speed up the etch rate of the film \cite{TiNr}. We hypothesize that the multi-modal etching behaviour observed for the TiN film could result from this mechanism. This interpretation is also supported~\cite{supplementary} by our estimates of the active surface from SEM imaging as well as \gls{AFMy} performed in the cross-sectional area presented in Figure \ref{Fig3}.

\section*{Conclusions}
The etching method presented here is particularly well suited for applications requiring high selectivity, where the underlying substrates or intermediate layers must remain intact during patterning to preserve their properties and ensure device functionality. The Supplementary Information shows an example of how we employed this process as the finishing step in an hybrid dry+wet etching for the nanopatterning of \gls{Nb}/\gls{Si} weak-link structures whose functionality could not tolerate substrate damage. Another relevant application is in the fabrication of high-quality coplanar waveguide (CPW) resonators, where \gls{TiN} membranes are patterned over through-silicon vias (TSV) electrode layers \cite{HighQ}. In such devices, maintaining the integrity of the electrode is essential, as any damage can significantly degrade resonator performance and a hybrid or fully wet etching approach can offer a clear advantage. Furthermore, in ferroelectric capacitor devices, such as TiN/HfZrO$_{\text{x}}$/TiN stacks \cite{Cap}, an \gls{SC-1} wet etching process can be beneficial for patterning the \gls{TiN} electrode while preserving the ferroelectric properties of the HfZrO$_{\text{x}}$ layer, which are closely tied to its crystallographic structure and overall quality.

However, one well-known challenge of wet etching approaches is the overetching of metal under the mask, resulting from the isotropic nature of the process; the Supplementary Information includes a quantitative impact assessment for our case. Finally, the operational lifetime of the SC-1 solution may be constrained by the H$_{2}$O$_{2}$ thermal decomposition, impacting solution stability and etching results unless appropriate process control measures are adopted.

Nevertheless, this process offers significant mitigation to specific challenges in dry etch techniques, such as enhanced selectivity of metal compared to Si and SiO$_{2}$ and reducing substrate damage. In addition, its experimental set up and performance makes it practically suitable for microfabrication, considering, for example, its ability to etch both Nb and TiN thin films with thicknesses of order 100 nm within minutes, enabling rapid processing times and precise etch-stop control. Another important aspect is the simplicity of the process, the moderately low solution temperatures, and lack of hazardous reactants or generation of dangerous byproducts: enhancing safety, accessibility, and responsible manufacturing.

Future efforts will focus on optimizing the SC-1 wet etching recipe to characterize its process stability and reproducibility, as well as comparing its performance against conventional dry etching processes in terms of sidewall profile, pattern fidelity, surface roughness, and selectivity with other mask or substrate materials.

\section*{Acknowledgements}
We acknowledge Karl-Magnus Persson and Olli-Pekka Kilpi for the insightful discussions on the potential applications of this process. We thank also Enni Hartikainen and Stefan Mertin for their assistance on the sputtering of the TiN films, as well to Manika Maharjan, Sari Ahlfors, and Elina Haustola for their cleanroom fabrication support.

\section*{Conflict of interest}
The authors have no conflicts of interest to disclose.

\section*{Funding}
This work was supported by the European Union’s HORIZON-RIA programme (Grant 101135240-JOGATE, managed by Chips JU) and by Research Council of Finland through project No. 356542-SUPSI.

\section*{Data availability}
The data that support the findings of this study are available from the corresponding author upon reasonable request.

\bibliography{References}

\clearpage
\appendix








\renewcommand{\thesection}{S\arabic{section}}
\renewcommand{\thetable}{S\arabic{table}}
\renewcommand{\thefigure}{S\arabic{figure}}
\setcounter{figure}{0}
\setcounter{page}{1}

\clearpage

\section*{Supplementary Information for: SC-1 Etching of Niobium and Titanium Nitride Thin Films}

\section{Example of challenging selectivity in a dry etching \mbox{process}}
\label{Supplementarysection: Selectivity}

\begin{figure}[ht]
  \centering
  \includegraphics[scale=1.2]{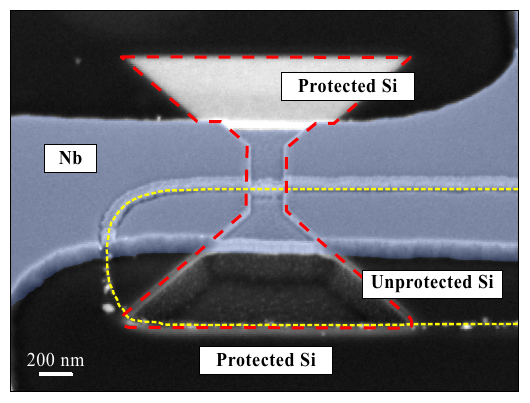}
  \caption{Superconducting Nb weak link half-patterned using reactive ion etching.}
  \label{FigS1}
\end{figure}

Reactive ion etching was tested for the patterning of superconducting Nb weak links, aimed to investigate proximity effect in Josephson junctions (JJs). 
Figure \ref{FigS1} shows a process-development device, composed of a \mbox{20 nm} aluminium oxide layer protecting a substrate except for a rectangular opening (yellow-dotted), and a 100 nm thick Nb strip. Here, the
rectangular opening is vertically misaligned with the Nb metallization strip due to error in lithography.

During the electron beam lithography (EBL) patterning (bowtie shape) of a trench in the Nb strip, we partially etched the metal layer for approximately 50 nm. Noticeably, the exposed Si area is deeply etched by the plasma, resulting in a large pit. This reveals a strong etching selectivity issue: if we aim to fully clear out the Nb within the defined trench without affecting at all the Si underneath, impossibly precise control on the etch time would be required.

\clearpage

\section{Original SEM image from Figure 1}
\label{supplementarysection:SEMFig1}

\begin{figure}[ht]
  \centering
  \includegraphics[scale=1.2]{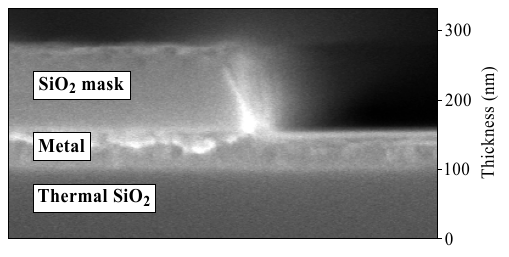}
  \caption{Cross-sectional \gls{SEM} image of a chip at the edge of a mask opening prior to wet etching, showing the initial layer stack (no false-color).}
  \label{FigS2}
\end{figure}

\section{Original SEM images from Figure 3}
\label{supplementarysection:SEMFig3}

\begin{figure}[ht]
  \centering
  \includegraphics[scale=0.83]{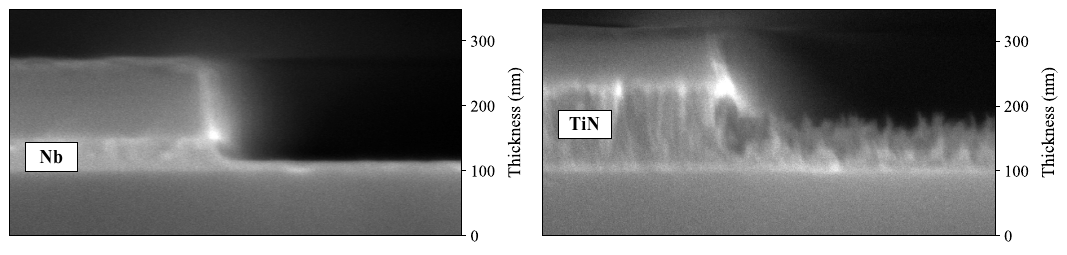}
  \caption{Cross-sectional  SEM  images  of the etched area from the Nb and TiN films when the etching process is approximately half-way:  after 80 s and 150 s for Nb and TiN, respectively (no false-color).}
  \label{FigS3}
\end{figure}

\clearpage
\section{Effective surface estimation and atomic force microscopy from the TiN film after 150 s of etching}
\label{supplementarysection:EffectiveArea}

\begin{figure}[ht]
  \centering
  \includegraphics[scale=.9]{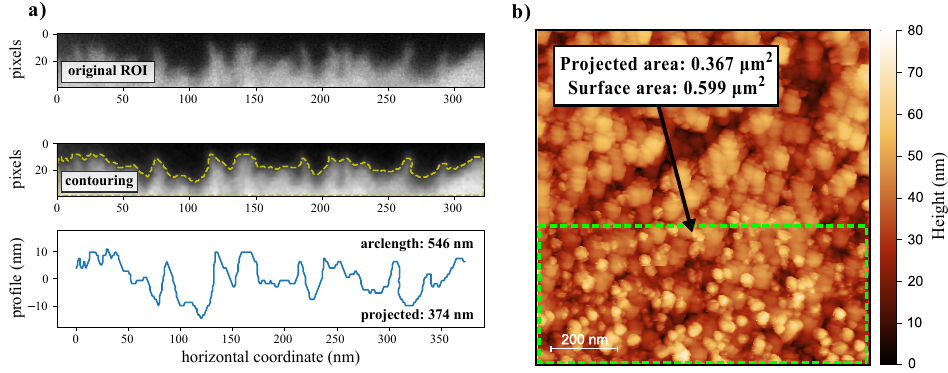}
  \caption{Effective surface a) profile and b) area estimation on the SEM imaged region presented in Figure 3, corresponding to the TiN film after 150 s of SC-1 etching.}
  \label{FigS4}
\end{figure}

The TiN effective surface profile after 150 s of etching, obtained from the cross-sectional SEM image presented in Figure 3, was estimated using an image processing approach that involved a threshold filtering and contour measurement algorithm from the Open Computer Vision library\footnote{Available at https://opencv.org}. Figure \ref{FigS4}a illustrates the SEM-imaged TiN profile, the resulting contour after applying the thresholding filter, and the final arclength estimation. Over a projected horizontal length of 374 nm, the estimated arclength profile extended to 546 nm, indicating a linear increase factor $\approx 1.46$. 

AFM was performed on the same sample, shown in Figure \ref{FigS4}b. Surface area was estimated on a specific region\footnote[1]{The AFM scanning acquisition exhibited difficulties due to potential residues being attached to the tip or tip damaging, hence, only a region with consistent topography and minimal artifacts was chosen for the estimation (green-dotted area in Figure \ref{FigS4}b).} within the scanned area and compared to its corresponding projected area. Results show a relative increase of $\approx 1.63$ in surface area, comparably with Figure \ref{FigS4}a. 

\section{Effect of mask undercut etching}
\label{supplementarysection:Undercut}

\begin{figure}[ht]
  \centering
  \includegraphics[scale=0.9]{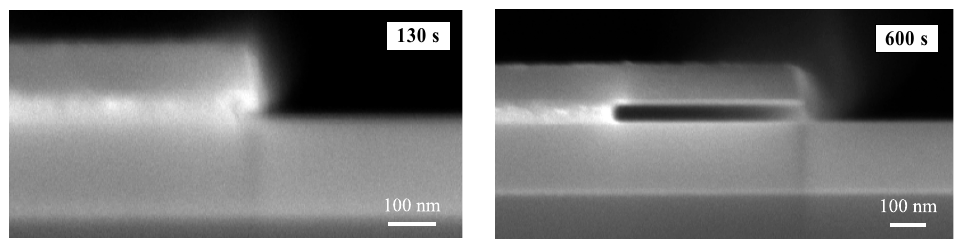}
  \caption{Effect of prolonged etching times on the Nb thin film samples, showcasing the development of a sidewall cavity after 130 s and 600 s of SC-1 etching.}
  \label{FigS5}
\end{figure}

Figure \ref{FigS5} illustrates the effect of mask undercut at prolonged etching times in the Nb thin film case. Mask undercut refers to the lateral etching beneath the mask layer, which occurs when the etching process continues beyond the intended duration (overetching). The left panel exemplifies the situation when 10 s of overetching (130 s in total) occurred, showing the development of a lateral cavity of approximately 50 nm. In contrast, the right panel shows the case when the etching time is extended to a total of 600 s, where the lateral cavity increases to approximately 500 nm. 

The effect of mask undercut etching can result in an increase in the linewidth on fabricated devices, deviating from the intended pattern design. 

\end{document}